\documentclass[fleqn,twoside]{article}
\usepackage{espcrc2}

\usepackage{graphicx}
\usepackage[figuresright]{rotating}

\newcommand{\AmS}{{\protect\the\textfont2
  A\kern-.1667em\lower.5ex\hbox{M}\kern-.125emS}}

\title{The KASCADE-Grande Experiment and the LOPES Project}

\author{A.F.~Badea\address[IKUK]{Institut f\"ur Kernphysik, Forschungszentrum
        Karlsruhe and
        Institut f\"ur Experimentelle Kernphysik, Universit\"at Karlsruhe,
	P.O. Box 3640, D-76021 Karlsruhe, Germany}\thanks{corresponding 
	author, e-mail: Florin.Badea@ik.fzk.de}\thanks{on leave of absence from 
	{\rm $^b$}},
        T.~Antoni\addressmark[IKUK], 
        W.D.~Apel\addressmark[IKUK], 
        K.~Bekk\addressmark[IKUK], 
        A.~Bercuci\address[NIPNE]{National Institute of Physics and Nuclear 
	Engineering,
	P.O. Box Mg-6, RO-7690 Bucharest, Romania},
	M.~Bertaina\address[DFGU]{Dpt.di Fisica Generale dell'
	Universita,
	Via Pietro Giuria n.1, I-10125 Torino, Italy and
	Istituto di Fisica dello Spazio Interplanetario del CNR, sez. di
	Torino,
	C. so Fiume n. 4, I-10133 Torino, Italy},
        H.~Bl\"umer\addressmark[IKUK],
        H.~Bozdog\addressmark[IKUK],
        I.M.~Brancus\addressmark[NIPNE],
	M.~Br\"uggemann\address[USIG]{Universit\"at Siegen, Fachbereich Physik,
	Universit\"at Siegen, Emmy-Noether-Campus,
	Walter-Flex-Str.~3, D-57068 Siegen, Germany},
	P.~Bucholz\addressmark[USIG],
	A.~Chiavassa\addressmark[DFGU],
        K.~Daumiller\addressmark[IKUK],
	F.~di Pierro\addressmark[DFGU],
        P.~Doll\addressmark[IKUK],
	R.~Engel\addressmark[IKUK],
        J.~Engler\addressmark[IKUK], 
	H.~Falcke\address[ASTRON]{ASTRON, 7990 AA Dwingeloo, The Netherlands},
        F.~Fe{\ss}ler\addressmark[IKUK],
	P.L.~Ghia\addressmark[DFGU],
        H.J.~Gils\addressmark[IKUK],
        R.~Glasstetter\address[UW]{Fachbereich Physik, Universit\"at Wuppertal,
	42097 Wuppertal, Germany},
        A.~Haungs\addressmark[IKUK], 
        D.~Heck\addressmark[IKUK],
        J.R.~H\"orandel\addressmark[IKUK],
	A.~Horneffer\address[MPIB]{Max-Planck-Institut f\"ur Radioastronomie,
	D-53121 Bonn, Germany},
	T.~Huege\addressmark[MPIB],
        K.-H.~Kampert\addressmark[UW],
	G.W.~Kant\addressmark[ASTRON],
        H.O.~Klages\addressmark[IKUK],
	Y.~Kolotaev\addressmark[USIG],
        G.~Maier\addressmark[IKUK],
        H.J.~Mathes\addressmark[IKUK],
        H.J.~Mayer\addressmark[IKUK], 
        J.~Milke\addressmark[IKUK],
	C.~Morello\addressmark[DFGU], 
        M.~M\"uller\addressmark[IKUK],
	G.~Navarra\addressmark[DFGU], 
        R.~Obenland\addressmark[IKUK],
        J.~Oehlschl\"ager\addressmark[IKUK],
        S.~Ostapchenko\addressmark[IKUK]\thanks{on leave of absence from Moscow 
	State University, Russia},
        M.~Petcu\addressmark[NIPNE],
	S.~Plewnia\addressmark[IKUK],
        H.~Rebel\addressmark[IKUK],
        A.~Risse\address[SINS]{Soltan Institute for Nuclear Studies,
	PL-90950 Lodz, Poland}, 
        M.~Roth\addressmark[IKUK], 
        H.~Schieler\addressmark[IKUK], 
        J.~Scholz\addressmark[IKUK],
	M.~St\"umpert\addressmark[IKUK], 
        T.~Thouw\addressmark[IKUK],
	G.~Toma\addressmark[NIPNE],
	G.C.~Trinchiero\addressmark[DFGU],
        H.~Ulrich\addressmark[IKUK],
	S.~Valchierotti\addressmark[DFGU],
	J.~van Buren\addressmark[IKUK],
	C.M.~de Vos\addressmark[ASTRON],
	W.~Walkowiak\addressmark[USIG],
        A.~Weindl\addressmark[IKUK], 
        J.~Wochele\addressmark[IKUK], 
        J.~Zabierowski\addressmark[SINS],
	S.~Zagromski\addressmark[IKUK],
	D.~Zimmermann\addressmark[USIG]
       }
       
\begin{document}

\begin{abstract}
KASCADE-Grande is the extension of the multi-detector setup KASCADE to cover 
a primary cosmic ray energy range from \mbox{100 TeV} to \mbox{1 EeV}. 
The enlarged EAS experiment provides comprehensive observations of cosmic rays 
in the energy region around the knee. Grande is an array of 
\mbox{700 x 700 m$^2$} equipped with 37 plastic scintillator stations 
sensitive to measure energy deposits and arrival times of air shower particles. 
LOPES is a small radio antenna array to operate in conjunction with 
KASCADE-Grande in order to calibrate the radio emission from cosmic ray air showers. 
Status and capabilities of the KASCADE-Grande experiment and the LOPES project 
are presented.
\vspace*{-0.5cm}
\vspace{1pc}
\end{abstract}

\maketitle

\section{Introduction}

\begin{figure}[h]
\vspace*{-0.4cm}
\includegraphics[width=7.5cm]{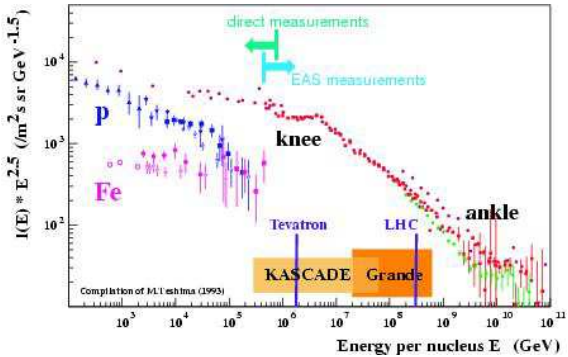}
\vspace*{-1.3cm}
\caption{Primary cosmic ray flux and primary energy range covered by 
KASCADE-Grande.}
\vspace*{-0.5cm}
\label{knee}
\end{figure}

KASCADE-Grande allows a full coverage of the energy range around the so called 
"knee" of the primary cosmic ray spectrum (see Fig.~\ref{knee}). There are different 
theoretical attempts to explain the mystery of the origin of the "knee" called 
change of the slope in the all-particle energy spectrum of cosmic rays. 
Either the knee energy is defined by the probability of an escape from the 
magnetic field 
of our Galaxy or by the limit of acceleration in supernova remnants or 
other galactic objects. Both 
classes of theories (diffusion or acceleration based) predict knee positions
occurring at constant rigidity of the particles. 
On the other hand, the hypothesis of new hadronic
interaction mechanisms at the knee energy, as for example the production of
heavy particles in $pp$ collisions, implies an atomic mass 
dependence of the knee positions. 
It is obvious that only detailed measurements
and analysis of the primary energy spectra for the different incoming 
particle types can validate or disprove some of 
these models (see also \cite{Kampe04}). 
From KASCADE~\cite{Anton03} measurements we do know that at a few times 
\mbox{10$^{15}$ eV} the knee is due to light elements~\cite{Anton02}, that the 
knee positions depend on the kind of the incoming particle and that cosmic rays 
around the knee arrive our Earth isotropically~\cite{Gmaier1,Gmaier2}. 
KASCADE-Grande~\cite{Navar04,Haung03}, measuring higher energies, will 
prove, if exists, the knee corresponding to heavy elements.
Additionally KASCADE could show that no hadronic interaction model 
describes very well 
cosmic ray measurements in the energy range of the knee and 
above~\cite{Ulric04}. These model uncertainties are
due to the lack of accelerator data at these energies and especially for the 
forward direction of collisions. Multi-detector systems like 
KASCADE and KASCADE-Grande offer the possibility
of testing and tuning the different hadronic interaction models.
With its capabilities KASCADE-Grande is also the testbed for the 
development and calibration of new air-shower
detection techniques like the measurement of EAS radio emission.

\section{The KASCADE Experiment}

\begin{figure}[h]
\vspace*{-0.4cm}
\includegraphics[width=7.5cm]{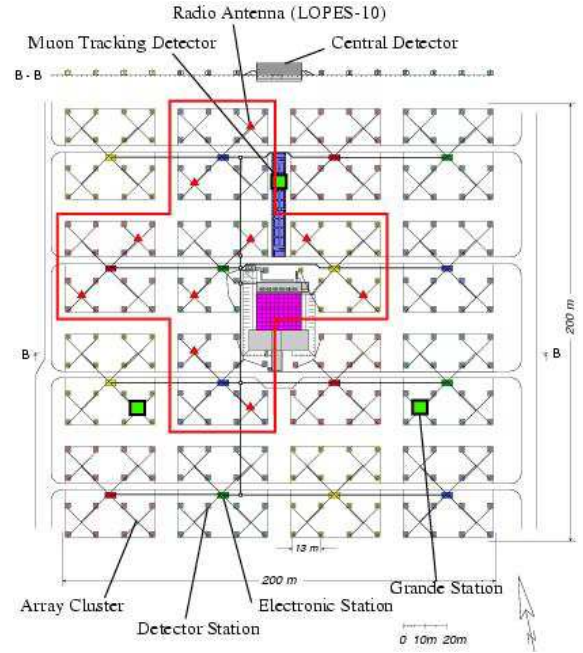}
\vspace*{-1.3cm}
\caption{The main detector components of the KASCADE experiment: (the 16
clusters of) Field Array, Muon Tracking Detector and Central Detector. The location of 
10 radio antennas is also displayed, as well as three stations of the Grande array.}
\vspace*{-0.5cm}
\label{KASCADE_radio}
\end{figure}

The KASCADE experiment (Fig.~\ref{KASCADE_radio}) measures 
showers in a primary energy range from \mbox{100 TeV} to \mbox{80 PeV} 
and provides 
multi-parameter measurements on a large number of observables concerning 
electrons, muons at 4 energy thresholds, and hadrons.
The main detector components of KASCADE are the Field Array, Central
Detector and Muon Tracking Detector. The Field Array measures the 
electromagnetic and 
muonic
components with \mbox{5 MeV} and \mbox{230 MeV} energy threshold, respectively.
It provides basic information about the arrival direction and core position as 
well as number of muons and electrons of the observed shower. The Muon Tracking Detector
measures the angles-of-incidence of muons with \mbox{800 MeV} energy threshold.
The Central Detector consists mainly of a hadron sampling calorimeter; 
three other
components - trigger plane, MWPC, LST - offer additional valuable information on the 
penetrating muonic component at \mbox{490 MeV} and \mbox{2.4 GeV} energy 
thresholds. Main results of KASCADE are summarized in \cite{Kampe04}.

\section{The KASCADE-Grande Experiment}

The multi-detector concept of the KASCADE experiment has been translated to 
higher primary energies through KASCADE-Grande to solve the threefold problem: 
unknown primary energy, primary mass and characteristics of the hadronic interactions. 
This requires measurements on many shower parameters by using a 
multi-detector system to get redundant informations. Consistent experimental 
and simulated data are compared in order to estimate the mass and energy of the 
primary particles. Then multidimensional simulated 
distributions of observables can be compared with experimental ones in order to 
validate the interaction models.

\subsection{The Grande Array}

\begin{figure}[ht]
\vspace*{-0.01cm}
\includegraphics[width=7.5cm]{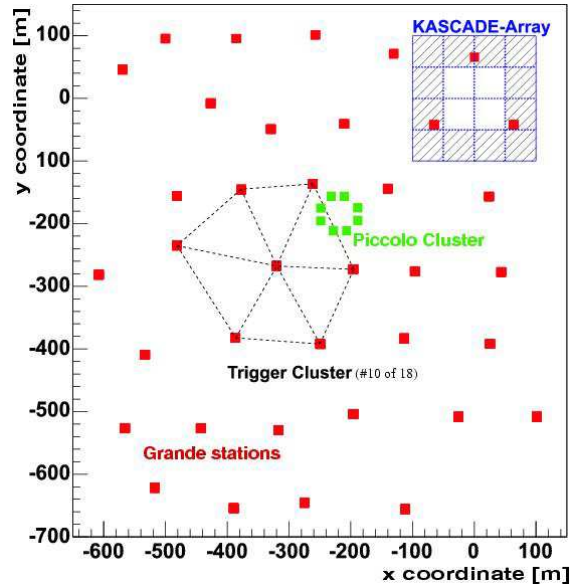}
\vspace*{-1.3cm}
\caption{Sketch of the KASCADE-Grande experiment.}
\vspace*{-0.5cm}
\label{grande}
\end{figure}

The 37 stations of the Grande Array (Fig.~\ref{grande}), located inside 
Karlsruhe Research Center, 
extend the cosmic ray measurements up to primary 
energies of \mbox{1 EeV}. 
The Grande stations, \mbox{10 m$^2$} of plastic scintillator detectors each, are 
spaced at approximative \mbox{130 m} 
covering a total area of \mbox{$\sim$ 0.5 km$^2$}. There are 16 scintillator 
sheets in a station read-out by 16 high gain photomultipliers; 4 of the 
scintillators are read-out also by 4 low gain PMs. The covered dynamic range is 
up to \mbox{3000 mips/m$^2$}. 
A trigger signal is build when 7 stations in a hexagon (trigger cluster, see 
Fig.~\ref{grande}) are fired. Therefore the Grande Array consists of 18 
hexagons with a total trigger rate of \mbox{0.5 Hz}.

\subsection{The Piccolo Array}

Additionally to the Grande Array a compact array, named Piccolo, has 
been build in order to provide a fast trigger to KASCADE ensuring joint 
measurements for showers with cores located far from the KASCADE array.
The Piccolo array consists of 8 stations with \mbox{11 m$^2$} plastic 
scintillator
each, distributed over an area of \mbox{360 m$^2$}. One station contains 12 
plastic scintillators organized in 6 modules; 3 modules form a so-called 
electronic station providing ADC and TDC signals. A Piccolo trigger is built
and sent to KASCADE and Grande
when at least 7 out of the 48 modules of Piccolo are fired. Such a logical 
condition leads to a trigger rate of \mbox{0.3 Hz}.

\section{Measurements at KASCADE-Grande}

\begin{figure}[t]
\vspace*{0.01cm}
\centering
\includegraphics[width=7.3cm]{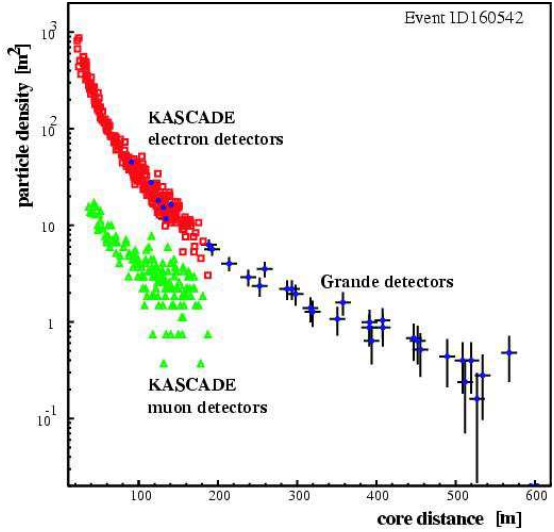}
\vspace*{-1.0cm}
\caption{Particle densities in the different detector types of KASCADE-Grande 
measured for a single event.}
\vspace*{-0.5cm}
\label{LDFexample}
\end{figure}

Fig.~\ref{LDFexample} shows, for a single event,
the lateral distribution of electrons and muons reconstructed with KASCADE and 
the charge particle densities 
measured by the Grande stations. This example shows the capabilities of
KASCADE-Grande and the high quality of the data. The 
KASCADE-Grande reconstruction procedure follows iterative steps: shower 
core position, 
angle-of-incidence and total number of charged particles are estimated from 
Grande Array data; the muon densities and with that the reconstruction of the
total muon number is provided by KASCADE muon detectors; by subtracting it from 
the number of charged particles, the total electron number is estimated.  The 
reconstruction accuracy of the
\begin{figure}[h]
\vspace*{0.01cm}
\centering
\includegraphics[width=7.3cm]{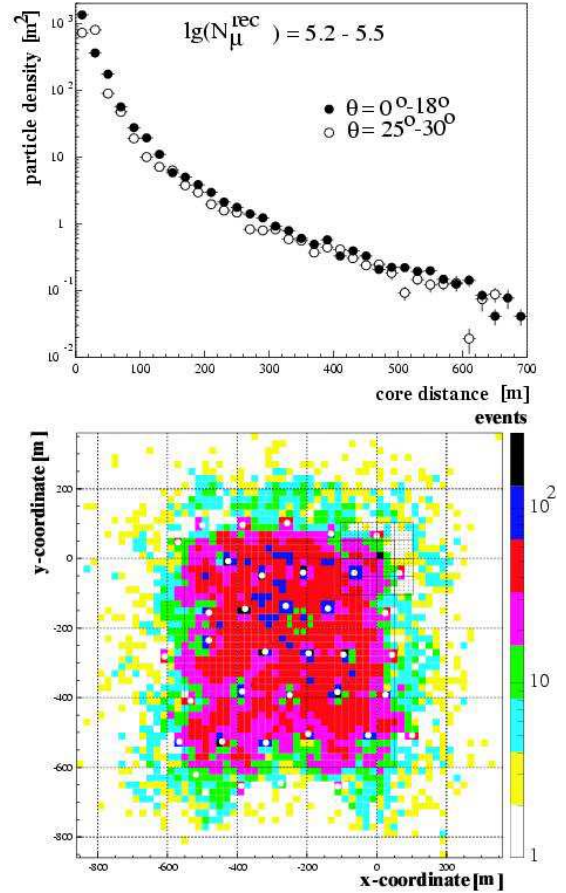}
\vspace*{-.5cm}
\caption{Mean lateral distributions of charged particles for two zenith angles
ranges and distribution of the shower core positions 
reconstructed from Grande Array measurements at a 3-day test-run.}
\label{test-run}
\vspace*{-0.5cm}
\end{figure}
shower core position and direction is in the order of \mbox{4 m} (\mbox{13 m})
and $0.18^{\circ}$ ($0.32^{\circ}$) with $68$\% ($95$\%) confidence level for
\begin{figure}[h]
\includegraphics[width=7.5cm]{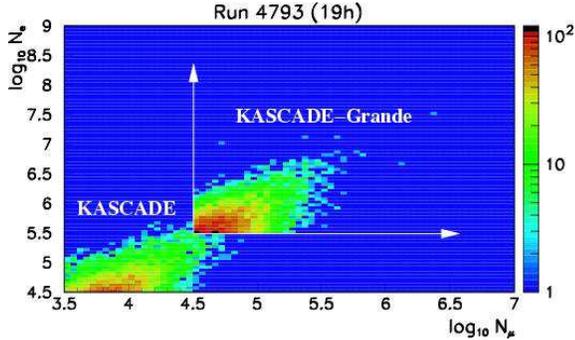}
\vspace*{-1.3cm}
\caption{Comparison between KASCADE and KASCADE-Grande data for a combined 
test-run.}
\vspace*{-0.5cm}
\label{april04_data}
\end{figure}
simulated proton and iron showers at \mbox{100 PeV} primary energy and 
$22^{\circ}$ zenith angle~\cite{Glass03}. The statistical uncertainty of the
shower sizes are around $15$\% for both, the total numbers of electrons and 
muons. The critical point of the KASCADE-Grande reconstruction is the 
estimation of the muon number due to the limited sampling of the muon lateral 
distribution by the KASCADE muon detectors.
The systematic uncertainty for the muon number depends on the radial range of 
the data measured by the KASCADE array and the chosen lateral distribution
function~\cite{Glass03}. 
In Fig.~\ref{test-run} 
the mean lateral distributions of charged particles for two zenith angle ranges 
and the shower core distribution for a 3-day test-run are presented.
The data of this test-run are used to optimize the reconstruction
procedures at KASCADE-Grande.
Both distributions show reasonable results and are the basis for further
reconstruction improvements. 

At the KASCADE experiment, the two-dimensional distribution 
shower size - truncated
number of muons played the fundamental role in reconstruction of energy spectra 
of single mass groups. For the same run time, due to its 10 times 
larger area compared with KASCADE, the Grande Array sees a significant number 
of showers at primary energies \mbox{$\sim$10 times} higher 
(Fig.~\ref{april04_data}). Hence fig.~\ref{april04_data} shows the capability of
KASCADE-Grande to perform an unfolding procedure like in 
KASCADE~\cite{Kampe04}.

To improve further the data quality a self-triggering, dead-time free 
FADC-based DAQ system will be implemented in 
order to record the full time evolution of energy deposits in the Grande 
stations at an effective sampling rate of \mbox{250 MHz} and high resolution of 
12 bits in two gain ranges~\cite{Andre04}. 
This will lead to an intrinsic electron-muon separation of the data signal at
the Grande Array. 

\section{The LOPES Project}

LOFAR ({\bf LO}far {\bf F}requency {\bf AR}ray) will be a new digital 
interferometer and is an attempt to revitalize astrophysical research at 
\mbox{20-200 MHz} with the means of modern technology~\cite{LOFAR04}. 
The received waves from outer space will be digitized and sent to a central 
super-cluster of computers. 
LOFAR combines
the advantages of a low gain antenna (large field of view) and high-gain antenna
(high sensitivity and background suppression); these makes it a powerful tool also
for studying radio emission in air-showers. A "{\bf LO}far {\bf P}rototyp{\bf E} 
{\bf S}tation" (LOPES) is under construction at the KASCADE-Grande location in
order to test the LOFAR technology and demonstrate its capability for EAS
measurements.

The interaction of the shower electrons and positrons with the Earth magnetic
field leading to geosynchrotron emission is expected to be the dominant factor 
for radio emission at the cosmic ray air shower development~\cite{Falck03}. 
Due to the low attenuation of radio waves, the emission
is primarily a measure of the total electron and positron content in the shower
maximum. The shower thickness is of order of the wavelength in the 
\mbox{100 MHz} regime, thus coherence and interference effects are important 
at this frequency band. As a consequence, the shower geometry is imprinted in 
the measured wavefront of the radio antennas. Again, coherence of the wave front
would lead to a quadratic increase of the radio pulse with primary particle 
energy~\cite{Verno68}.
\begin{figure}[ht]
\vspace*{-0.01cm}
\centering
\includegraphics[width=7.cm]{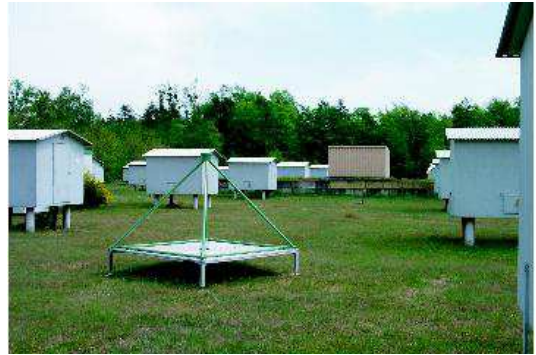}
\vspace*{-0.8cm}
\caption{Dipole radio antenna in the Field Array of KASCADE. A Grande station
on top of the Muon Tracking Detector is also seen.}
\label{radio-kg}
\vspace*{-0.5cm}
\end{figure}
For ultra-high energies radio detection is expected to have the most favorable 
signal-to-noise ratio compared to all other forms of secondary 
radiation~\cite{Gorha02}.
\begin{figure}[ht]
\vspace*{-0.01cm}
\centering
\includegraphics[width=7.cm]{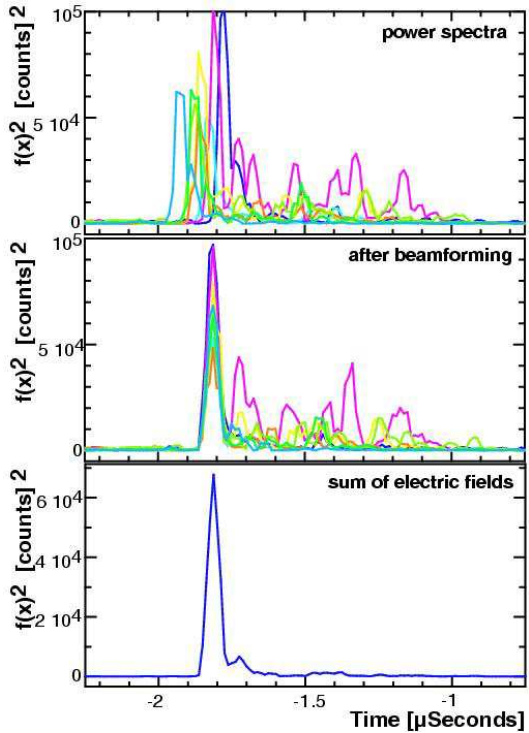}
\vspace*{-0.8cm}
\caption{Steps of the reconstruction of a strong air shower 
event~\cite{Horne04} detected by LOPES. In the upper part the received powers 
(squares of the electric fields) of 8 antennas before 
beamforming are overlayed. Powers after time shifting using 
the KASCADE shower direction information are displayed in the second panel.
A clear signal is proved by displaying the power after beamforming of the
electric field (lower part).}
\label{radio_raw}
\vspace*{-0.5cm}
\end{figure}
Also, in contrast to optical methods, the technique allows round the clock 
observations, and as radio has low attenuation, will allow large 
detector volumes. 

At present, LOPES operates 10 dipole radio antennas (see Fig.~\ref{radio-kg}) 
in coincidence with KASCADE~\cite{Horne02}.
The antennas are positioned in 5 out of the 16 clusters of KASCADE, 2 of them 
per cluster (see Fig.~\ref{KASCADE_radio}). The radio data is collected when a 
trigger is received from the KASCADE array. The logical condition for 
trigger is at least 10 out of the 16 clusters to be fired. This translates to 
primary energies above \mbox{$10^{16}$ eV}; such showers are detected at a rate
of 2 per minute. The antennas operate in the frequency range of \mbox{40-80 MHz}.
A preliminary analysis of the first data has already been 
performed~\cite{Horne04}. 
Fig.~\ref{radio_raw} shows a particularly bright event as an example. 
A crucial element of the detection method is the digital beamforming which 
allows to place a narrow antenna beam in the direction of the cosmic ray event.
This is possible because the phase information of the radio waves is preserved 
by the digital receiver and the cosmic ray produces a coherent pulse. 
This method is also very effective in suppressing interference from the particle 
detectors which all radiate incoherently.
 
In the near future, 30 antennas will be installed at KASCADE-Grande.
The FADC system planned for the Grande stations may play a key role in 
deconvoluting
and subtracting the radio signals produced by the particle detectors.
With an amount of 1000 measured events above \mbox{$10^{17}$ eV}, KASCADE-Grande
will be able to calibrate the radio emission in extensive air showers with
high accuracy.

\section{Conclusions}

The~extension~of~KASCADE~to~the~KASCADE-Grande experiment, accessing higher 
primary energies, will prove the existence of a knee-like structure  
corresponding to heavy elements. KASCADE-Grande will keep the multi-detector 
concept for tuning different interaction models at 
primary energies up to \mbox{$10^{18}$ eV}. There are promising perspectives 
for detecting radio emission in extensive air showers~with the LOPES set up 
and to perform a calibration of the radio pulses. This can lead to a 
possible new detection tool for future experiments.


\begin{thebibliography}{9}
\bibitem{Kampe04} K.H. Kampert et al. - KASCADE collab., these proceedings.
\bibitem{Anton03} T. Antoni et al. - KASCADE collab., Nucl. Instr. Meth. A 513
                 (2003) 429.
\bibitem{Anton02} T. Antoni et al. - KASCADE collab., Astrop. Phys. 16 (2002) 373.
\bibitem{Gmaier1} T. Antoni et al. - KASCADE collab., Astrophys. J. 604 (2004)
                  687.
\bibitem{Gmaier2} T. Antoni et al. - KASCADE collab., Astrophys. J. 608 (2004)
                  865.
\bibitem{Navar04} G. Navarra et al. - KASCADE-Grande collab., 
		  Nucl. Instr. Meth. A 518 (2004) 207.
\bibitem{Haung03} A. Haungs et al. - KASCADE-Grande collab., 
                  Proc. of 28$^{th}$ ICRC, Tsukuba, Japan (2003) p.985.
\bibitem{Ulric04} H. Ulrich, report Forschungszentrum Karlsruhe, Germany (2004) 
		  FZKA 6952 (in German).
\bibitem{Glass03} R. Glasstetter et al. - KASCADE-Grande collab., 
                  Proc. of 28$^{th}$ ICRC, Tsukuba, Japan (2003) p.781.
\bibitem{Andre04} W. Walkowiak et al. - KASCADE-Grande collab., "A FADC-based Data
                  Acquisition System for the KASCADE-Grande Experiment", Proc. of
		  IEEE Nuclear Science Symposium, Rome 2004, in preparation. 
\bibitem{LOFAR04} http://www.lofar.org/
\bibitem{Falck03} H. Falcke and P. Gorham, Astropart. Phys. 19 (2003) 477.
\bibitem{Verno68} S.N. Vernov et al., Can. J. Phys. 46 (1968) 241.
\bibitem{Gorha02} P. Gorham at al., Nucl. Instr. and Meth. A. 476 (2002) 476.
\bibitem{Horne02} A. Horneffer et al., 
                  Proc. of Proc. of 28$^{th}$ ICRC, Tsukuba, Japan (2003) p.969. 
\bibitem{Horne04} A. Horneffer et al.,
                  "LOPES - Detecting Radio Emission from Cosmic Ray Air 
		  Showers", proc. of SPIE conference, Glasgow (2004), in preparation.
\end{thebibliography}
\end{document}